\documentclass[aps,pra,twocolumn,superscriptaddress]{revtex4}
\usepackage{amssymb}
\usepackage{amsmath}
\usepackage{graphicx}
\usepackage{epsfig}
\usepackage{subfigure}
\usepackage{xcolor}

\begin{document}

\title{Rotation sensing in two coupled whispering-gallery-mode resonators with loss and gain}
\author{Tian Tian}
\affiliation{School of Science, Changchun University, Changchun 130022, China}
\affiliation{Key Laboratory of Materials Design and Quantum Simulation, Changchun University, Changchun 130022, China}
\author{Zhihai Wang}
\email{wangzh761@nenu.edu.cn}
\affiliation{Center for Quantum Sciences and School of Physics, Northeast Normal University, Changchun 130024, China}
\affiliation{Center for Advanced Optoelectronic Functional Materials Research, and Key Laboratory for UV-Emitting Materials and Technology of
Ministry of Education, Northeast Normal University, Changchun 130024, China}
\author{L. J. Song}
\email{ccdxslj@126.com}
\affiliation{Jilin Engineering Normal University, Changchun 130052, China}
\affiliation{Key Laboratory of Materials Design and Quantum Simulation, Changchun University, Changchun 130022, China}

\begin{abstract}
We theoretically propose a scheme to realize rotation sensing based on two coupled whispering-gallery-mode resonators {with loss and gain. We consider that the active resonator with gain is rotated while the passive one with loss is stationary. The rotation will induce Sagnac effect and we show that the eigenfrequencies of the supermodes are sensitive to the Sagnac-Fizeau shift.} Therefore, we can measure the average photon number in the steady state or the fluctuation spectrum of the output fields to detect the angular velocity of the rotation. We hope that our investigation will be useful in the {design} of quantum gyroscope based on spinning resonators.
\end{abstract}

\maketitle
\section{Introduction}
Quantum gyroscope, which plays an important role in inertial navigation, is aiming to perform rotation sensing {with high precision} by use of quantum effects. Up to now, {the studies about quantum gyroscope have covered many various quantum systems}, such as photon and matter-wave interferometers~\cite{dowling, dowling1,SA}, cold atom system~\cite{cold1,cold2,cold3}, solid spin system~\cite{spin1,spin2,spin3,spin4,spin5} , optomechanical system~\cite{opto1,opto2} and so on.

Whispering-gallery-mode (WGM) resonators have become versatile {platforms for both of fundamental physical researches and technology applications.} For example, people have demonstrated the non-Hermitian physics associated with parity-time ($\mathcal{PT}$) symmetry phase transition~\cite{peng,chang} and proposed its application in low-threshold lasers~\cite{jing,feng}, quantum metrology~\cite{liu} and quantum sensing for nano-particles~\cite{chen,chen1}. Recently, {the WGM resonator with mechanical rotation (also named as spinning resonator)~\cite{SM,jiang} is invoking more and more attentions}, including rotation enhanced quantum sensing~\cite{hui}, non-reciprocal photon blockade~\cite{Li,huang} and so on.

{Based on the above achievements, it is natural to investigate the performance of quantum gyroscope based on spinning WGM resonator. The underlying physics behind the spinning WGM resonator gyroscope is Sagnac effect~\cite{sa1,sa2}. That is, the frequency of the optical mode will be shifted by the rotation and the amount of the shift is linearly dependent on the angular velocity of the rotation}. As a result, the response of the system to the external driving or noise is sensitive to the angular velocity, and it supplies us an effective approach to perform rotation sensing.

{Based on the Sagnac effect, we propose a theoretical scheme of quantum gyroscope in two coupled WGM resonators system with loss and gain. We consider that the active resonator with gain is mechanically rotated while the passive one with loss is stationary. Here, the coupling between the two resonators will induce normal mode splitting and the eigenfrequencies of the supermodes are related to the angular velocity of the rotation.} We demonstrate that the angular velocity can be detected by measuring either the {average photon numbers in the steady state or the fluctuation spectra of the output fields of resonators}. Furthermore, we also compare the fluctuation spectra with and without gain, our results show that the optical gain will enhance the performance of rotation sensing {by narrowing and heightening the spectrum peaks. However, superior to the quantum sensing schemes based on $\mathcal{PT}$ symmetry phase transition, our scheme is not necessary to work in the parameter regime nearby the exceptional point.} Therefore, it is more convenient to be realized in experiments.

The rest of the paper is organized as follows. In Sec.~\ref{model}, we introduce the theoretical model and study the energy spectrum of the system, which is described by a non-Hermitian Hamiltonian. In Sec.~\ref{classical}, we study the average photon number in the active resonator. {In Sec.~\ref{quantum}, we investigate the fluctuation spectra of the output fields.} At last, we give a brief conclusion in Sec.~\ref{con}.

\section{Model}
\label{model}

\begin{figure}[tbp]
\begin{centering}
\includegraphics[width=1\columnwidth]{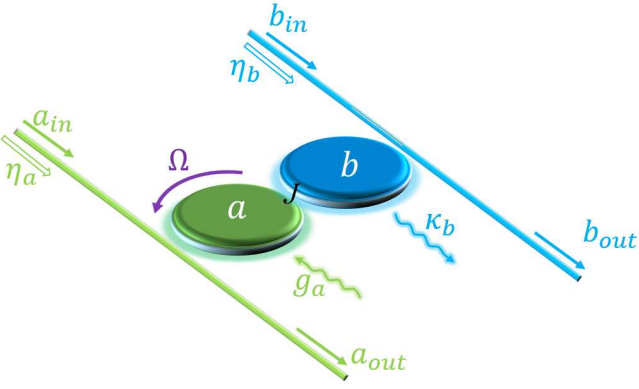}
\par\end{centering}
\caption{Schematic diagram of two coupled whispering-gallery-mode resonators, which are coherently driven by the classical fields. The resonator $a$ with gain rotates with the angular velocity $\Omega$ while the resonator $b$ with loss is stationary. }
\label{scheme}
\end{figure}

As shown in Fig.~\ref{scheme}, we study the system of two coupled WGM resonators $a$ and $b$, which are driven coherently by the external classical fields. We consider that $a$ is an active resonator {with gain}, and $b$ is a passive one {with loss}. As reported in Ref.~\cite{peng}, the gain in resonator $a$ can be realized by fabricating from $\rm{Er^{3+}}$-doped silica. Both of the two WGM resonators support two kinds of optical modes, which propagate along the clockwise (CW) direction and counterclockwise (CCW) direction, respectively. However, the classical fields only drive the CCW mode in resonator $a$ and the CW mode in resonator $b$. {Therefore, both of them are treated as single-mode resonator in our consideration.} The coupling strength $J$ between the two resonators can be tuned by controlling their distance. Furthermore, the active resonator is rotated counterclockwisely with the angular velocity $\Omega$, while the passive resonator and the waveguides are stationary. Recently, such kind of rotating system has been realized and used to demonstrate the optical nonreciprocity~\cite{SM}. In this paper, we will perform quantum sensing to the angular velocity by studying the average values of the steady state and {fluctuation spectra of the output fields}.

The Hamiltonian of the system can be written as $H=H_0+H_I$, where ($\hbar=1$)
\begin{eqnarray}
H_0&=&(\omega_{a}-\Delta+ig_a)a^{\dagger}a+(\omega_{b}-i\kappa_b)b^{\dagger}b
\nonumber \\&&+J(a^{\dagger}b+ab^{\dagger}),\label{H0}\\
H_I&=&(\eta_{a}a^{\dagger}e^{-i\omega_dt}+h.c.)+(\eta_{b}b^{\dagger}e^{-i\omega_dt}
+h.c.).
\end{eqnarray}
Here, $\omega_a$ and $\omega_b$ are the resonant frequencies of resonator $a$ and $b$, respectively. $\eta_a$ and $\eta_b$ are the driving strengths of the classical fields, {which possess the same} frequency $\omega_d$. In what follows, we will assume $\omega_a=\omega_b=\bar\omega$, and set $\eta_a$ and $\eta_b$ to be real numbers. $g_a(\kappa_b)$ is the total gain (decay) rate of resonator $a$ ($b$), whose detailed definition will be given below.  $\Delta$ in Eq.~(\ref{H0}) is the rotation induced Sagnac-Fizeau shift for resonator $a$, which is expressed as~\cite{GB}
\begin{equation}
\Delta=\frac{nR\Omega\omega_a}{c}(1-\frac{1}{n^2}-\frac{\lambda}{n}\frac{dn}{d\lambda})
\end{equation}
with $n$ and $R$ the refractive indices and radius of resonator $a$. $\lambda$ and $c$ are the wavelength and speed of light {in vacuum}, respectively.

\begin{figure}[tbp]
\begin{centering}
\includegraphics[width=1\columnwidth]{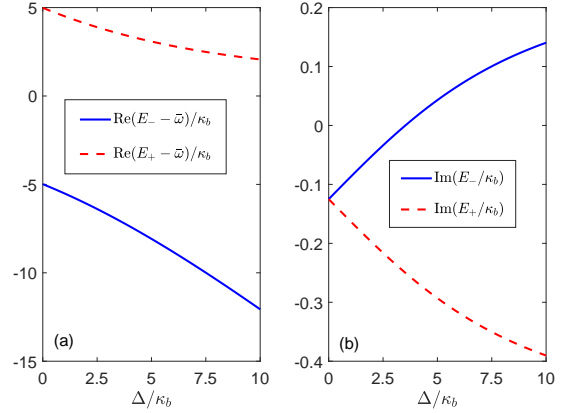}
\par\end{centering}
\caption{The real (a) and imaginary (b) parts of $E_\pm$. The parameters are set as $g_a=0.5\kappa_b, J=5\kappa_b$.}
\label{level}
\end{figure}

The non-Hermitian Hamiltonian $H_0$ can also be written as
\begin{equation}
H_{0}=\left(\begin{array}{cc}
a^{\dagger} & b^{\dagger}\end{array}\right)\mathcal{H}\left(\begin{array}{c}
a\\b\end{array}\right).
\end{equation}
where
\begin{equation}
\mathcal{H}=\left(\begin{array}{cc}
\bar\omega-\Delta+ig_{a} & J\\
J & \bar\omega\end{array}\right),
\label{fH}
\end{equation}
and its eigenvalues are
\begin{equation}
E_{\pm}=\frac{2\bar\omega-\Delta+ig_a\pm\sqrt{(\Delta-ig_a)^2+4J^2}}{2}.
\label{eigen}
\end{equation}
{It shows that} the coupling between the two resonators induces a normal mode splitting. {The real (imaginary) parts of the eigenvalues $E_\pm$ of $\mathcal{H}$ represent the frequencies (decay rates) of the two supermodes.} It is observed from Fig.~\ref{level}(a) that, the frequencies of the two supermodes are modified by the rotation. {Moreover, the decay rates of the two supermodes are also sensitive to the Sagnac-Fizeau shift $\Delta$ as shown in Fig.~\ref{level}(b).} This kind of rotation induced distortion has been discussed in Ref.~\cite{MJ}. {It also implies that we can detect the energy spectrum of the system to acquire the angular velocity of the rotation. A common approach to detect the energy spectrum of a quantum system is to drive it weakly and measure the response of the system to the driving field(s), and this is what we will discuss in the rest parts of this paper.}

The {dependence on the time of} the Hamiltonian can be eliminated by transferring into the rotating frame, then the Hamiltonian becomes
\begin{eqnarray}
H'&=&(\bar\Delta-\Delta+ig_a)a^{\dagger}a+(\bar\Delta-i\kappa_b)b^{\dagger}b
\nonumber \\&&+J(a^{\dagger}b+ab^{\dagger})+(\eta_aa^{\dagger}+\eta_b b^{\dagger}+h.c.),\label{Hr}
\end{eqnarray}
where $\bar\Delta:=\bar\omega-\omega_d$. According to the Heisenberg equations for the motion of the operators and adding the noise terms, we will obtain the quantum Langevin equations as
\begin{eqnarray}
\frac{d}{dt}a&=&	-[i(\bar\Delta-\Delta)-g_{a}/2)]a-iJb-i\eta_{a}\nonumber\\&&+\sqrt{\kappa_{ex,a}}a_{in}
+\sqrt{\kappa_{0,a}}a_{in}^{(0)}+\sqrt{g}a_{in}^{(g)},\label{L1}\\
\frac{d}{dt}b&=&	-(i\bar\Delta+\kappa_{b}/2)b-iJa-i\eta_{b}\nonumber \\ &&+\sqrt{\kappa_{ex,b}}b_{in}
+\sqrt{\kappa_{0,b}}b_{in}^{(0)}.\label{L2}
\end{eqnarray}
Here, $g$ is the gain rate of the resonator mode $a$ and $g_{a}=g-\kappa_{a}$. {The total decay rate of the resonator is described by $\kappa_{m}=\kappa_{ex,m}+\kappa_{0,m}\,(m=a,b)$. $\kappa_{ex,m}$ is the coupling rate of resonator $m$ to the corresponding waveguide
while $\kappa_{0,m}$ is its coupling rate to the surrounding environments.}

The zero-mean-noise operators associated with the loss of the resonators satisfy~\cite{Scully}
\begin{eqnarray}
\langle o_{in}(t) o_{in}^{\dagger}(\tau)\rangle=\delta(t-\tau),
\langle o_{in}^{\dagger}(t) o_{in}(\tau)\rangle=0, \\
\langle o_{in}^{(0)}(t) o_{in}^{(0)\dagger}(\tau)\rangle=\delta(t-\tau),
\langle o_{in}^{(0)\dagger}(t) o_{in}^{(0)}(\tau)\rangle=0,
\end{eqnarray}
for $o=a,b$. The noise operators associated with the gain of resonator $a$ obey~\cite{Scully,GS,BH,KV}
{\begin{equation}
\langle a_{in}^{(g)}(t) a_{in}^{(g)\dagger}(\tau)\rangle=0,
\langle a_{in}^{(g)\dagger}(t) a_{in}^{(g)}(\tau)\rangle=\delta(t-\tau).
\end{equation}}

In quantum mechanics, an arbitrary operator can be separated by its classical part (which is a $c$ number average value) and quantum fluctuation (which is still an operator). In other words, it is reasonable to write the operators $a$ and $b$ as
$a=\alpha+\delta a$ and $b=\beta+\delta b$. Here, $\alpha$ and $\beta$ are the average values,  $\delta a$ and $\delta b$ are the quantum fluctuations. {In Sec.~\ref{quantum}, we will use $a$ and $b$ as a shorthand for $\delta a$ and $\delta b$ since the discussion in this part deals exclusively with fluctuations.}

\section{Average values in the steady state}
\label{classical}

In this section, we will discuss the behavior of average values in the steady state. The {quantum fluctuations} will be studied in the next section.

It follows from Eqs.~(\ref{L1}) and (\ref{L2}) that the average values $\alpha$ and $\beta$ yield
\begin{eqnarray}
\frac{d\alpha}{dt}&=&-[i(\bar\Delta-\Delta)-g_{a}/2]\alpha-iJ\beta-i\eta_{a},\label{alpha1}\\
\frac{d\beta}{dt}&=&-(i\bar\Delta+\kappa_{b}/2)\beta-iJ\alpha-i\eta_{b}.\label{beta1}
\end{eqnarray}
The system is stable only if the imaginary parts of all the eigenvalues of the matrix $\mathcal{H}$ [defined in Eq.~(\ref{fH})] are negative. In what follows, we will choose the parameters inside the stable regime.

The average values in the steady state are solved as
\begin{eqnarray}
\alpha&=&\frac{(\bar\Delta-i\kappa_{b}/2)\eta_{a}-J\eta_{b}}
{J^{2}-(\bar\Delta-\Delta+ig_{a}/2)(\bar\Delta-i\kappa_{b}/2)},\label{alpha}\\
\beta&=&\frac{(\bar\Delta-\Delta+ig_{a}/2)\eta_{b}-J\eta_{a}}
{J^{2}-(\bar\Delta-\Delta+ig_{a}/2)(\bar\Delta-i\kappa_{b}/2)}\label{beta}.
\end{eqnarray}

\begin{figure}
\begin{centering}
\includegraphics[width=1\columnwidth]{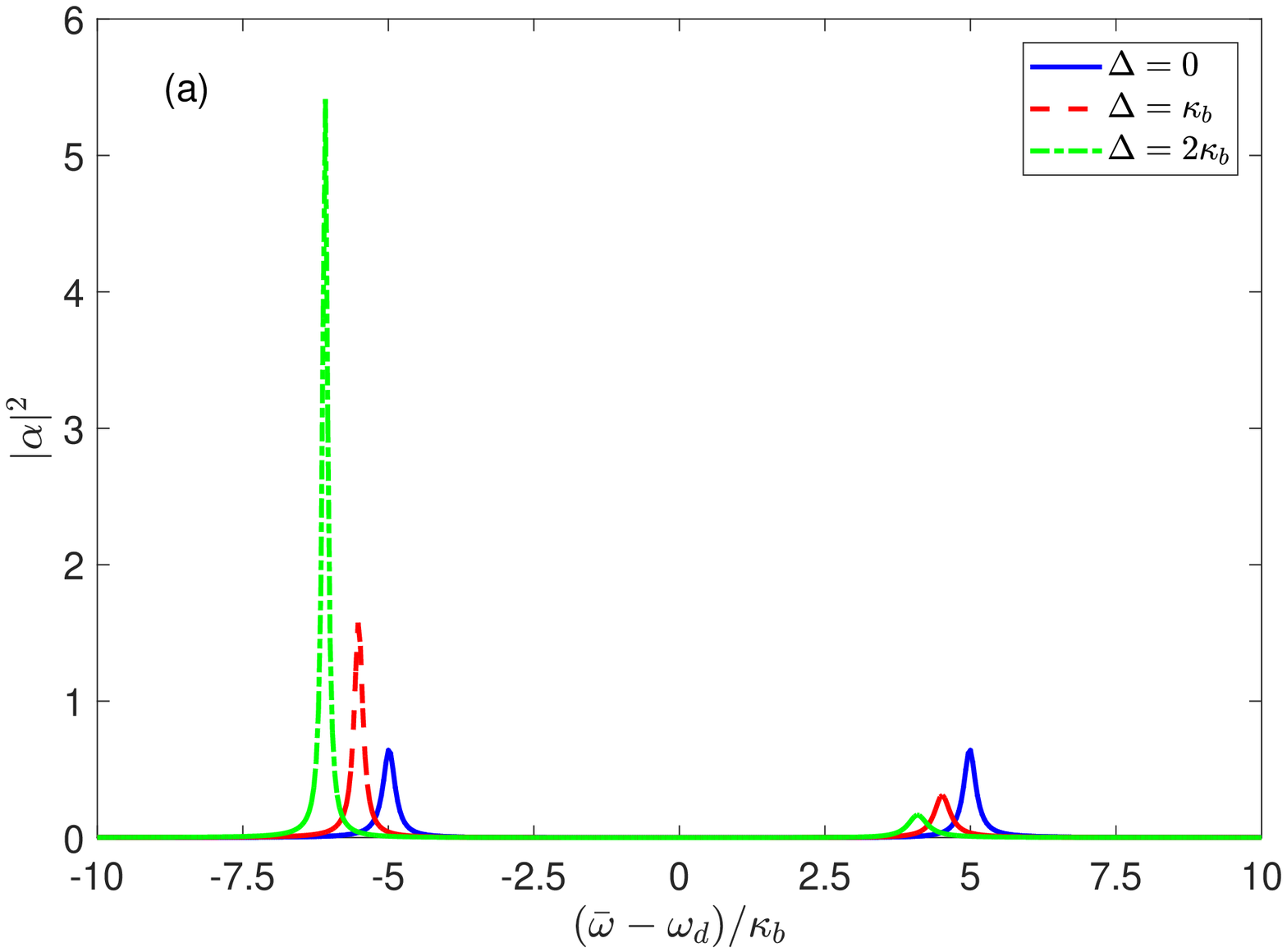}
\par\end{centering}
\begin{centering}
\includegraphics[width=1\columnwidth]{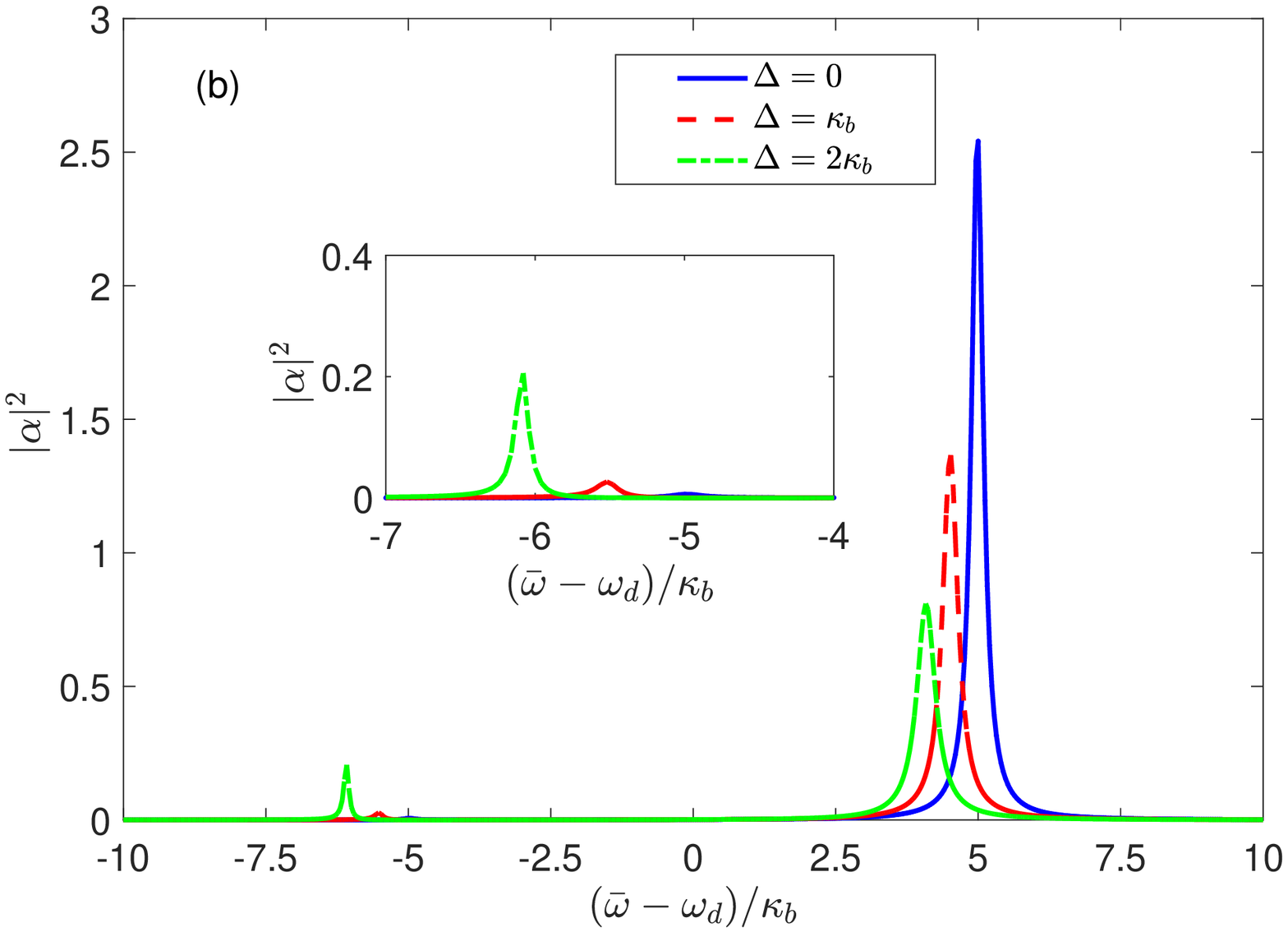}
\par\end{centering}
\caption{Average photon number $|\alpha|^2$ as a function of the frequency
of the driving field for different Sagnac-Fizeau shifts $\Delta$. The parameters are set as $g_a=0.5\kappa_b, J=5\kappa_b$ and (a) $\eta_a=0.2\kappa_b,\eta_b=0$, (b) $\eta_a=\eta_b=0.2\kappa_b$. {The inset in lower panel [panel (b)] shows the resolved interval for $(\bar\omega-\omega_d)/\kappa_b\in{(-7,-4)}$.} }
\label{average}
\end{figure}

To perform rotation sensing in our system, we will plot the behavior of average photon number $|\alpha|^2$ in resonator $a$ as a function of the frequency of the driving field(s), for different Sagnac-Fizeau shifts $\Delta$ in Fig.~\ref{average}. {We remark that the expression of $\beta$ in Eq.~(\ref{beta}) has a similar form to $\alpha$ in Eq.~(\ref{alpha}),  so the curves for $\beta$ will not be shown in our paper.}

We now consider the case when only the mode in resonator $a$ is classically driven, that is $\eta_b=0$. As shown in Fig.~\ref{average}(a), the average photon number shows two peaks, at which the driving field is resonant with the supermodes, this is {the normal mode splitting phenomenon. Compared with the right peak, the left peak is much narrower and the average photon number is much larger [as shown in Fig.~\ref{average}(a)]. This results is consistent with those shown in Fig.~\ref{level}(b) where the imaginary parts of $E_\pm$ are plotted to demonstrate the different lifetimes of the two supermodes.} Furthermore, we show that the positions and heights of the peaks are dependent on the Sagnac-Fizeau shift $\Delta$, which supplies us an effective approach to perform rotation sensing by measuring the average photon number in the steady state.

Surprisingly, when both of the resonators are driven by the classical fields with same intensity ($\eta_a=\eta_b=:\eta$), {the results in Fig.~\ref{average}(b) show} that the left peak nearly disappears while the right peak remains. This phenomenon can be explained by writing the Hamiltonian $H'$ [see Eq.(7)] in the supermode representation. To this end, we firstly consider a simple case without rotation, gain and loss. Then, the Hamiltonian can be written as
\begin{equation}
H'=(\bar\Delta+J)A^\dagger A+(\bar\Delta-J)B^\dagger B+\sqrt{2}\eta(A^\dagger+A),
\label{Hsuper}
\end{equation}
where the {annihilation operator of symmetric supermode is defined as $A:=(a+b)/\sqrt{2}$, while {that of the antisymmetric} supermode is defined as $B:=(a-b)/\sqrt{2}$. The last term in Eq.~(\ref{Hsuper}) only contains the terms of symmetric supermode $A$, it implies that the antisymmetric supermode $B$ is decoupled from the external driving fields, so that the left peak nearly disappears as shown in Fig.~\ref{average}(b)[Compare to Fig.~\ref{average}(a)]. Next}, when the rotation, gain and loss of the system are taken into account, the {symmetric and antisymmetric} nature of the two supermodes are broken. {In such a situation, both of the two supermodes are coupled to the classical driving fields, but the coupling intensity for supermode $B$ is much smaller {than} that for supermode $A$.} Therefore, as shown in the inset of Fig.~\ref{average}(b), there is also a small peak when the driving fields are resonant with the supermode $B$. In this sense, when the two WGM resonators are both driven with same intensity, we have two signatures to {verify} the rotation, one is the shift of the right photon number peak while the other one is the appearance of the left peak.

Let us go back to the case where only the resonator $a$ is driven, that is, $\eta_b=0$. the Hamiltonian without rotation, gain and loss can be written as
\begin{equation}
H'=(\bar\Delta+J)A^\dagger A+(\bar\Delta-J)B^\dagger B+\frac{\eta_a}{\sqrt{2}}(A^\dagger+A-B^\dagger-B).
\label{Hsuper1}
\end{equation}
{The lase term} implies that both of the supermodes are driven coherently with the same intensity. As a result, the average photon number behaves as a double-peak-type curve as shown in Fig.~\ref{average}(a). {Here, the asymmetry of the two peaks comes from different decay rates of the two supermodes.}

\section{Fluctuation spectrum}
\label{quantum}

{In the above section, we have studied the response of the coupled WGM resonators system to the external driving field{s}. It shows that the average photon number in the steady state is sensitive to the rotation of the active resonator, so it is possible to serve as a rotation sensor.} In this section, we will take the quantum noise into consideration and investigate the fluctuation spectra of the output fields.

In what follows, we will use the notation $a$ and $b$ to denote the fluctuation operators without any confusion. According to Eqs.~(\ref{L1},\ref{L2},\ref{alpha1},\ref{beta1}), the fluctuations satisfy
\begin{eqnarray}
\frac{d}{dt}a&=&-[i(\bar\Delta-\Delta)-g_{a}/2]a-iJb\nonumber \\&&
+\sqrt{\kappa_{ex,a}}a_{in}+\sqrt{\kappa_{0,a}}a_{in}^{(0)}+\sqrt{g}a_{in}^{(g)},\\
\frac{d}{dt}b&=&-(i\bar\Delta+\kappa_{b}/2)b-iJa\nonumber \\&&+\sqrt{\kappa_{ex,b}}b_{in}+\sqrt{\kappa_{0,b}}b_{in}^{(0)}.
\end{eqnarray}
By introducing the Fourier transformation of operators
\begin{equation}
o(\omega)=\int_{-\infty}^{\infty}o(t)e^{i\omega t}dt,\, o^{\dagger}(\omega)=\int_{-\infty}^{\infty}o^{\dagger}(t)e^{i\omega t}dt,
\end{equation}
the operators for the resonator modes are solved as
\begin{eqnarray}
a(\omega)&=&\frac{-iJ\mathcal{B}_{in}(\omega)+iF_b(\omega)
\mathcal{A}_{in}(\omega)}{N(\omega)},\\
b(\omega)&=&\frac{-iJ\mathcal{A}_{in}(\omega)+iF_a(\omega)
\mathcal{B}_{in}(\omega)}{N(\omega)},
\end{eqnarray}
where
\begin{subequations}
\begin{eqnarray}
N(\omega)&=&J^{2}-F_a(\omega)F_b(\omega), \\
F_a(\omega)&=&\bar\Delta-\Delta-\omega+ig_{a}/2,\\
F_b(\omega)&=&\bar\Delta-\omega-i\kappa_{b}/2,\\
\mathcal{A}_{in}(\omega)&=&\sqrt{\kappa_{ex,a}}{a}_{in}(\omega)
+\sqrt{\kappa_{0,a}}{a}_{in}^{(0)}(\omega)+\sqrt{g}{a}_{in}^{(g)}(\omega),\nonumber \\ \\
\mathcal{B}_{in}(\omega)&=&\sqrt{\kappa_{ex,b}}{b}_{in}(\omega)
+\sqrt{\kappa_{0,b}}{b}_{in}^{(0)}(\omega).
\end{eqnarray}
\end{subequations}

Besides, the correlation relation for the noise {operators} in the frequency domain can be obtained as
\begin{eqnarray}
\langle a_{in}^{\dagger}(\omega)a_{in}(\Omega)\rangle&=&0,\,\langle a_{in}(\omega)a_{in}^{\dagger}(\Omega)\rangle=2\pi\delta(\omega+\Omega),\nonumber\\
\langle a_{in}^{(0)\dagger}(\omega)a_{in}^{(0)}(\Omega)\rangle&=&0,\,\langle a_{in}^{(0)}(\omega)a_{in}^{(0)\dagger}(\Omega)\rangle=2\pi\delta(\omega+\Omega),\nonumber\\
\langle b_{in}^{\dagger}(\omega)b_{in}(\Omega)\rangle&=&0,\,\langle b_{in}(\omega)b_{in}^{\dagger}(\Omega)\rangle=2\pi\delta(\omega+\Omega),\nonumber \\
\langle b_{in}^{(0)\dagger}(\omega)b_{in}^{(0)}(\Omega)\rangle&=&0,\,\langle b_{in}^{(0)}(\omega)b_{in}^{(0)\dagger}(\Omega)\rangle=2\pi\delta(\omega+\Omega),\nonumber\\
\langle a_{in}^{(g)}(\omega)a_{in}^{(g)\dagger}(\Omega)\rangle&=&0,\,\langle a_{in}^{(g)\dagger}(\omega)a_{in}^{(g)}(\Omega)\rangle=2\pi\delta(\omega+\Omega).\nonumber\\
 \end{eqnarray}

{The fluctuation spectrum is interesting in our coupled WGM resonators system, and it is experimentally measurable by the homodyne measurement of the output fields~\cite{hom}. By use of the input-output relation $a_{out}={a}_{in}+\sqrt{\kappa_{ex,a}}{a}$~\cite{DF} and the relation   $o^\dagger(\omega)=[o(-\omega)]^\dagger$, the fluctuation spectrum of the active resonator $a$  can be expressed as~\cite{DM,AM,JC}
\begin{eqnarray}
S_{a}(\omega)&=&\frac{1}{2}\int\frac{d\Omega}{2\pi}\langle a_{out}(\omega)a_{out}^{\dagger}(\Omega)+a_{out}^{\dagger}
(\Omega)a_{out}(\omega)\rangle\nonumber\\
&=&S_{a}^{(1)}(\omega)+S_{a}^{(2)}(\omega)+S_{a}^{(3)}(\omega),
\label{sa}
\end{eqnarray}
where
\begin{subequations}
\begin{eqnarray}
S_{a}^{(1)}(\omega)&=&\frac{|N(\omega)+i\kappa_{ex,a}F_{b}(\omega)|^{2}
+\kappa_{0,a}\kappa_{ex,a}
|F_{b}(\omega)|^{2}}{2|N(\omega)|^{2}},\nonumber \\ \\
S_{a}^{(2)}(\omega)&=&\frac{g\kappa_{ex,a}
|F_{b}(\omega)|^{2}}{2|N(\omega)|^{2}},\\
S_{a}^{(3)}(\omega)&=&\frac{J^{2}\kappa_{ex,a}(\kappa_{ex,b}+\kappa_{0,b})}{2|N(\omega)|^{2}}.
\end{eqnarray}
\end{subequations}
 Here, the first term $S_a^{(1)}(\omega)$ comes from the contribution of the loss of resonator $a$, including the coupling to the waveguide and surrounding environments. The second term $S_a^{(2)}(\omega)$ comes from the contribution of the gain of resonator $a$. Due to the coupling between the resonators, these two terms are also related to the loss of resonator $b$.  The last term $S_a^{(3)}(\omega)$ is completely contributed by the inter-resonator coupling.}

\begin{figure}
\begin{centering}
\includegraphics[width=1\columnwidth]{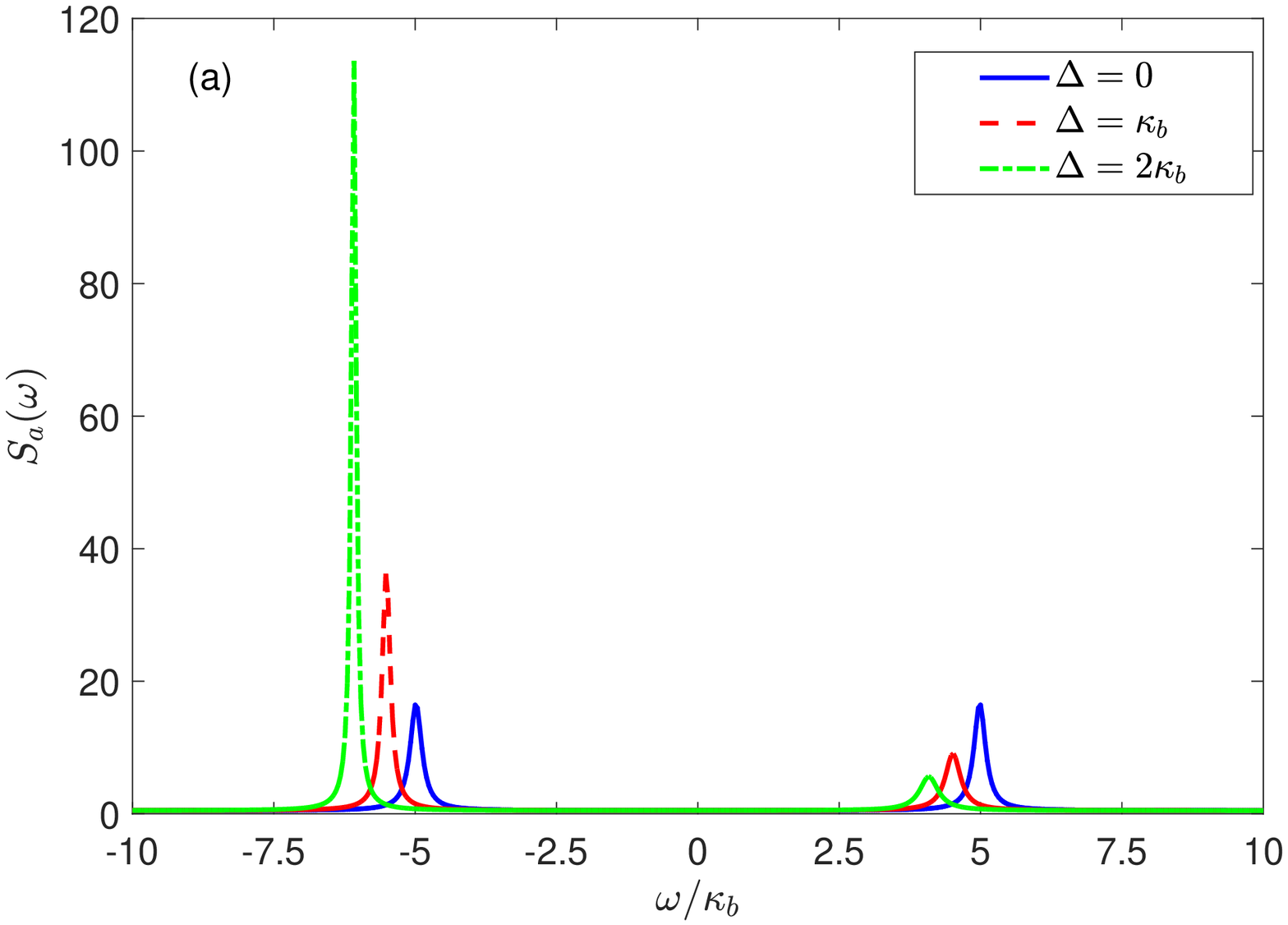}
\par\end{centering}
\begin{centering}
\includegraphics[width=1\columnwidth]{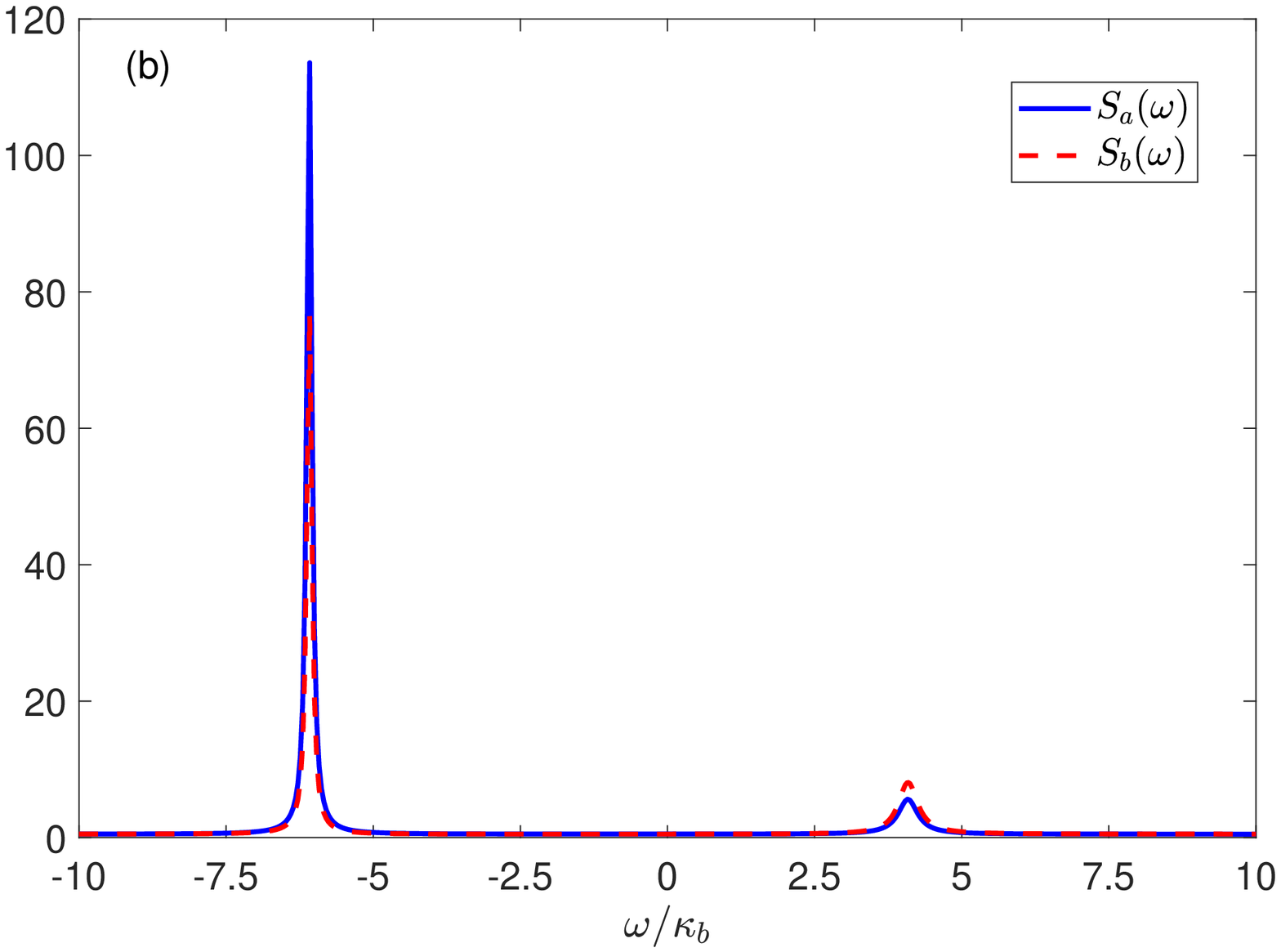}
\par\end{centering}
\begin{centering}
\includegraphics[width=1\columnwidth]{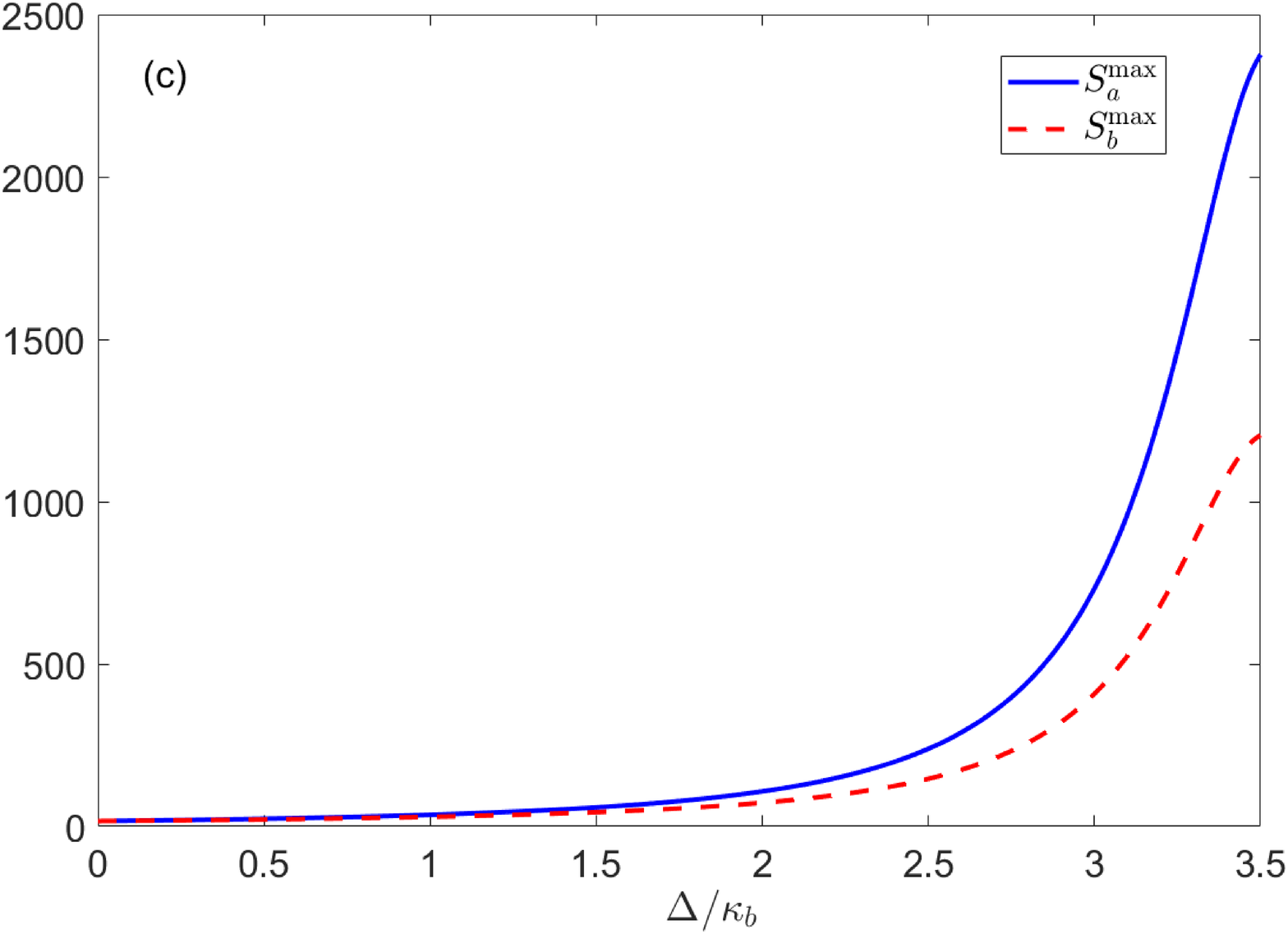}
\par\end{centering}
\caption{{(a) The fluctuation spectrum for the active resonator $a$ for different Sagnac-Fizeau shifts $\Delta$. (b) The comparison of the fluctuation spectra for active resonator $a$ and passive resonator $b$ for $\Delta/\kappa_b=2$. (c) The maximum values of the fluctuation spectra of the two resonators. The parameters are set as $g_a=0.5\kappa_b,\kappa_{ex,a}=\kappa_{0,a}=\kappa_{ex,b}=\kappa_{0,b}=0.5\kappa_b,  \omega_d=\bar\omega, J=5\kappa_b$.}}
\label{fluctuationspectrum}
\end{figure}

In Fig.~\ref{fluctuationspectrum}(a), we plot the fluctuation spectrum $S_a(\omega)$ when both of the resonators are driven resonantly ($\omega_d=\bar\omega$). Here, the active resonator $a$ experiences the noise associated with both of the gain and loss while the passive resonator $b$ experiences the noise associated with only the loss. {Therefore, the driving induced by the noise is out of balance for the two resonators, and the fluctuation spectrum behaves as a two-peak profile. That is, when the frequency is resonant with the two supermodes, the fluctuation spectrum will achieve its local maximum values. This behavior is similar to the average values $|\alpha|^2$ when only the mode in resonator $a$ is driven [as shown in Fig.~\ref{average}(a)].}

{Following the same procedure, we can obtain the expression for the fluctuation spectrum of resonator $b$ as
\begin{equation}
S_b(\omega)=S_b^{(1)}(\omega)+S_b^{(2)}(\omega)
\label{sb}
\end{equation}
where
\begin{subequations}
\begin{eqnarray}
S_b^{(1)}(\omega)&=&\frac{|N(\omega)+i\kappa_{ex,b}F_{a}(\omega)|^{2}
+\kappa_{0,b}\kappa_{ex,b}
|F_{a}(\omega)|^{2}}{2|N(\omega)|^{2}},\nonumber \\ \\
S_b^{(2)}(\omega)&=&\frac{J^{2}\kappa_{ex,b}(\kappa_{ex,a}
+\kappa_{0,a}+g)}{2|N(\omega)|^{2}}.
\end{eqnarray}
\end{subequations}
The first term $S_b^{(1)}(\omega)$ is the contribution from the coupling to the waveguide and the environment, while the second term $S_b^{(2)}(\omega)$ comes from the coupling to resonator $a$. In Fig.~\ref{fluctuationspectrum}(b), we plot $S_a(\omega)$ and $S_b(\omega)$  for a fixed Sagnac-Fizeau shift $\Delta=2\kappa_b$. It shows that $S_b(\omega)$ also possesses a double-peak structure. However, due to the fact that resonator $b$ is a passive one without gain, the value of the fluctuation spectrum is different from  that of resonator $a$ and the difference is much more apparent in the regime nearby the two peaks than other regimes.}

{The results in Figs.~\ref{fluctuationspectrum}(a) and (b) show that the distortion and the shift of the peak in the fluctuation spectra can serve as the indicator of the rotation. Furthermore, we note that the maximum values of the fluctuation spectra appear at the left peak. According to the eigenfrequencies of the supermodes given in Eq.~(\ref{eigen}), the left peak locates at the frequency $\omega_l=-\Delta/2-\sqrt{\Delta^2+4J^2}/2$ (That is, the real part of $E_-$. Here, we have replaced $\bar{\omega}$ by $\bar{\omega}-\omega_d$ and set it to be zero by considering a resonantly driving situation). In what follows, we denote $S_a(\omega_l)$ and $S_b(\omega_l)$ as $S_a^{\rm max}$ and $S_b^{\rm max}$ respectively, and plot them as functions of Sagnac-Fizeau shift $\Delta$ in Fig.~\ref{fluctuationspectrum}(c). From the expressions in Eqs.~(\ref{sa}-\ref{sb}), it seems that $S_a^{\rm max}$ and $S_b^{\rm max}$ are non-monotonic functions of $\Delta$. However, we restrict ourselves in the stable regime where the imaginary parts of all the eigenvalues of the matrix $\mathcal{H}$ [defined in Eq.~(\ref{fH})] are negative, and Fig.~\ref{fluctuationspectrum}(c) shows that both of $S_a^{\rm max}$ and $S_b^{\rm max}$  monotonously increase as the shift $\Delta$ becomes larger and $S_b^{\rm max}$ is smaller than $S_a^{\rm max}$.}

{The above results show that we can measure the fluctuation spectra (by for example the homodyne measurement) to acquire the information about the angular velocity of rotation. On one hand, when the resonators are driven resonantly, the peaks of the spectra locate at the eigenfrequencies of the supermodes, which are monotonously dependent on the Sagnac-Fizeau shift $\Delta$ by $E_\pm=\pm(\Delta+\sqrt{\Delta^2+4J^2})/2$. On the other hand, the results in Fig.~\ref{fluctuationspectrum}(c) show that the maximum values of the fluctuation spectra are also monotonous functions of $\Delta$ in the stable regime, so the angular velocity of the rotation can be determined by measuring the maximum values.}

\begin{figure}
\begin{centering}
\includegraphics[width=1\columnwidth]{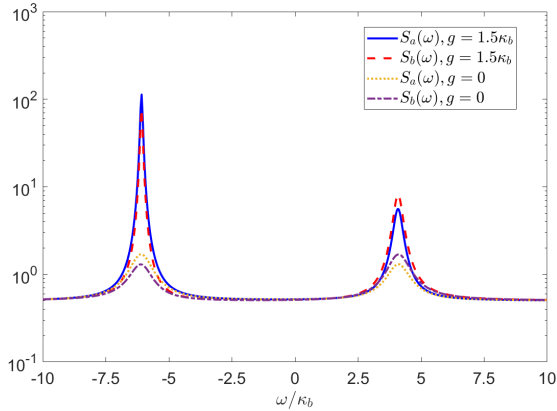}
\par\end{centering}
\caption{{The fluctuation spectra of the output fields of the resonator $a$ and $b$ with gain ($g=1.5\kappa_b$) and without gain ($g=0$). The other parameters are set to be same as Fig.~\ref{fluctuationspectrum}(b).}}
\label{compare}
\end{figure}

At last, we compare the results of the {fluctuation spectra $S_a(\omega)$ and $S_b(\omega)$ with gain ($g=1.5\kappa_b$) and without gain ($g=0$) for $\Delta=2\kappa_b$}. As shown in Fig.~\ref{compare}, the system with resonator gain will contribute two narrower and higher spectrum peaks, and is therefore
more sensitive in detecting the rotations.

\section{Conclusion}
\label{con}

Recently, the rotating WGM resonator has been realized experimentally, where the angular velocity has achieved by $\Omega\approx6.6$ kHz~\cite{SM}, and {a more faster rotation with frequency of GHz is also reported in the nano-rotor setup~\cite{JA,RR}}. In such a situation, the Sagnac-Fizeau shift is in the order of MHz to GHz, and therefore our theoretical proposal upon rotation sensing can be realized in the forthcoming experiments. {In some previous quantum sensing schemes with the assistance of loss and gain resonators, the sensitivity can usually be enhanced by working nearby the exceptional point~\cite{liu,chen,chen1}, where the $\mathcal{PT}$ symmetry phase transition happens. However, this restriction is not necessary in our theoretical proposal, therefore our scheme is more convenient to be realized experimentally.}

In conclusion, we have theoretically proposed a rotation sensing scheme based on two coupled WGM resonators with loss and gain. {The inter-resonator coupling induces a normal mode splitting in our scheme, and the eigenfrequencies of the supermodes are dependent on the rotation velocity. Originating from such dependence, we show that the angular velocity of the rotation can be sensed by measuring either the average values of steady state or the fluctuation spectra of the output fields.}  Moreover, the strength of the fluctuation spectrum with gain is much stronger than that without gain. We hope our studies can be useful for designing the active resonator based quantum gyroscopes.

\begin{acknowledgments}
 This work is supported by Jilin Province Science and Technology Development Plan (Grant Nos. 20180520175JH and 20170204023GX); Jilin Provincial Special Industry Innovation (Grant No. 2019C025);  Educational Commission of Jilin Province of China (Grant No. JJKH20190266KJ); The project of cultivating young teachers in Changchun University (Grant No. ZK201809) and National Natural Science Foundation of China (under Grants No. 11875011, No. 21703014).
\end{acknowledgments}


\begin{thebibliography}{99}

\bibitem{dowling}J. P. Dowling, Phys. Rev. A {\bf 57}, 4736 (1998).

\bibitem{dowling1}M. O. Scully and J. P. Dowling, Phys. Rev. A {\bf 48}, 3186 (1993).

\bibitem{SA}S. A. Haine, Phys. Rev. Lett. {\bf 116}, 230404 (2016).

\bibitem{cold1}S. Stringari, Phys. Rev. Lett. {\bf 86}, 4725  (2001).

\bibitem{cold2}I. Dutta, D. Savoie, B. Fang, B. Venon, C. L. G. Alzar, R. Geiger, and A. Landragin, Phys. Rev. Lett. {\bf 116}, 183003 (2016).

\bibitem{cold3}A. Gauguet, B. Canuel, T. L\'{e}v\`{e}que, W. Chaibi, and A. Landragin, Phys. Rev. A {\bf 80}, 063604 (2009).

\bibitem{spin1}T. W. Kornack, R. K. Ghosh, and M. V. Romalis, Phys. Rev. Lett. {\bf 95}, 230801 (2005).

\bibitem{spin2}A. Ajoy and P. Cappellaro, Phys. Rev. A {\bf 86}, 062104 (2012).

\bibitem{spin3}M. P. Ledbetter, K. Jensen, R. Fischer, A. Jarmola, and D. Budker,
Phys. Rev. A {\bf 86}, 052116 (2012).

\bibitem{spin4}J.-C. Jaskula, K. Saha, A. Ajoy, D. J. Twitchen, M. Markham, and P. Cappellaro, Phys. Rev. Appl. {\bf 11}, 054010 (2019).

\bibitem{spin5}A. A. Wood, E. Lilette, Y. Y. Fein, V. S. Perunicic, L. C. L. Hollenberg, R. E. Scholten, and A. M. Martin, Nature Phys. {\bf 13}, 1070 (2017).

\bibitem{opto1}K. Li, S. Davuluri, and Y. Li, Sci. China Phys. Mech. Astron. {\bf 61}, 90311 (2018).

\bibitem{opto2}S. Davuluri, K. Li, and Y. Li, New J. Phys. {\bf 19}, 113004 (2017).


\bibitem{peng}B. Peng, \c{S}. K. \"{O}zdemir, F. Lei, F. Monifi, M. Gianfreda, G. L. Long, S. Fan, F. Nori, C. M. Bender, and L. Yang,  Nat. Phys. {\bf 10}, 394 (2014).

\bibitem{chang}L. Chang, X. Jiang, S. Hua, C. Yang, J. Wen, L. Jiang, G. Li, G. Wang, and M. Xiao, Nat. Photon. {\bf 8}, 524 (2014).

\bibitem{jing}H. Jing, \c{S}. K. \"{O}zdemir, X.-Y. L\"{u}, J. Zhang, L. Yang, and F. Nori, Phys. Rev. Lett. {\bf 113}, 053604 (2014).

\bibitem{feng}L. Feng, Z. J. Wong, R.-M. Ma, Y. Wang, and X. Zhang, Science {\bf 346}, 972(2014).

\bibitem{liu}Z.-P. Liu, J. Zhang, \c{S}. K. \"{O}zdemir, B. Peng,  H. Jing,
X.-Y. L\"{u}, C.-W. Li, L. Yang, F. Nori, and Y.-x. Liu, Phys. Rev. Lett. {\bf 117},
110802 (2016).

\bibitem{chen}W. Chen, J. Zhang, B. Peng, \c{S}. K. \"{O}zdemir, X. Fan, and L. Yang, Photonics Research {\bf 6}, A23 (2018).

\bibitem{chen1}W. Chen, K. \"{O}zdemir, G. Zhao, J. Wiersig, and L. Yang, Nature {\bf 548}, 192 (2017).


\bibitem{SM}S. Maayani, R. Dahan, Y. Kligerman, E. Moses, A. U. Hassan, H. Jing,
F. Nori, D. N. Christodoulides, and T. Carmon,  Nature {\bf 558},
569 (2018).

\bibitem{jiang}Y. Jiang, S. Maayani, T. Carmon, F. Nori, and H. Jing, Phys. Rev. Appl. {\bf 10}, 064037 (2018).

\bibitem{hui}H. Jing, H. L\"{u}, \c{S}. K. \"{O}zdemir, T. Carmon, and F. Nori,
Optica {\bf 5}, 1425 (2018).


\bibitem{Li}B. Li, R. Huang, X. Xu, A. Miranowicz, and H. Jing, Photonics Research {\bf 7}, 630 (2019).

\bibitem{huang}R. Huang, A. Miranowicz, J.-Q. Liao, F. Nori, and H.
Jing, Phys. Rev. Lett. {\bf 121}, 153601 (2018).

\bibitem{sa1}E. J. Post, Rev. Mod. Phys. {\bf 39}, 475 (1967).

\bibitem{sa2}W.W. Chow, J. Gea-Banacloche, L. M. Pedrotti, V. E. Sanders,
W. Schleich, and M. O. Scully, Rev. Mod. Phys. {\bf 57}, 61 (1985).


\bibitem{GB}G. B. Malykin, Phys. Usp. {\bf 43}, 1229 (2000).

\bibitem{MJ}M. J. Grant, M. J. F. Digonnet, Proc. SPIE {\bf 10934},  109340T (2019).

\bibitem{Scully} M. O. Scully and M. S. Zubairy, \textit{Quantum optics} (Cambridge
university press, 1997).

\bibitem{GS}G. S. Agarwal and K. Qu, Phys. Rev. A {\bf 85}, 031802(R) (2012).

\bibitem{BH}B. He, L. Yang, Z. Zhang, and M. Xiao, Phys. Rev. A {\bf 91}, 033830
(2015).

\bibitem{KV} K. V. Kepesidis, T. J. Milburn, J. Huber, K. G. Makris, S. Rotter, and P. Rabl, New J. Phys. {\bf 18}, 095003 (2016).

\bibitem{DF}D. F. Walls and G. J. Milburn, \textit{Quantum Optics} (Springer, 2008).

\bibitem{hom}{V. Giovannetti and D. Vitali, Phys. Rev. A {\bf 63}, 023812 (2001).}

\bibitem{DM}D. Malz, L. D. T\'{o}th, N. R. Bernier, A. K. Feofanov, T. J.
Kippenberg, and A. Nunnenkamp,  Phys. Rev. Lett. {\bf 120}, 023601 (2018).

\bibitem{AM}A. Metelmann and A. A. Clerk, Phys. Rev. Lett. {\bf 112}, 133904
(2014).

\bibitem{JC}C. Jiang, L. N. Song, and Y. Li, Phys. Rev. A {\bf 97}, 053812 (2018).

\bibitem{JA}J. Ahn, Z. Xu, J. Bang, Y.-H. Deng, T. M. Hoang, Q. Han,
R.-M. Ma, T. Li, Phys. Rev. Lett. {\bf 121}, 033603 (2018).

\bibitem{RR}R. Reimann, M. Doderer, E. Hebestreit, R. Diehl, M. Frimmer,
D. Windey, F. Tebbenjohanns, and L. Novotny, Phys.
Rev. Lett. {\bf 121}, 033602 (2018).

\end{thebibliography}
\end{document}